\documentclass[aps,prl,twocolumn]{revtex4}
\usepackage{graphicx}
\usepackage{hyperref}
\input epsf
\begin{document}

\title{Ultra high densities of cold atoms in a holographically controlled dark SPOT trap}

\author{ N.\,Radwell*, G.\,Walker and \href{http://www.physics.gla.ac.uk/~sfrankea/}{S.\,Franke-Arnold}} 
\affiliation{SUPA, School of Physics and Astronomy, University of Glasgow, Glasgow G12 8QQ, UK}

\date{\today}

\begin{abstract} 

We demonstrate an atom trap geometry for $^{87}$Rb which is capable of producing ultra high atom densities. Reradiation forces, which usually limit high densities, can be avoided in dark spontaneous-force optical traps (dark SPOTs) by sheltering atoms from intense trapping light. Here we demonstrate a dynamic implementation of a dark SPOT, resulting in an increase in atom density by almost two orders of magnitude up to $1.3\times10^{12}\,$cm$^{-3}$. Holographic control of the trapping beams and dynamic switching between MOT and dark SPOT configuration allows us to optimise the trapping geometry. We have identified the ideal size of the dark core to be six times larger than the MOT. Our method also avoids unwanted heating so that we reach a record phase-space density for a MOT.

\end{abstract}

\maketitle

\emph{Introduction} -  High-density ultracold atomic vapors have emerged as promising candidates for quantum memories and high precision metrology \cite{sokolov10,lvovsky09}.  High atom densities furthermore  increase nonlinear and collectively enhanced effects, and are a prerequisit for the efficient loading of Bose condensates and dipole traps.  
Standard magneto-optical traps (MOTs) limit the obtainable densities for two distinct reasons: fistly, due to absorption the trapping light cannot penetrate a dense atomic sample so that cooling and confining processes cease to work efficiently; secondly, the interaction of the atoms with the trapping light ultimately forces the atoms apart through re-radiation processes, thereby increasing both size and temperature of the atom cloud \cite{walker90,weiner99}. A number of techniques have been employed to overcome these limitations. In compressed MOTs (CMOTs)  \cite{petrich94, depue00} the magnetic field is rapidly increased, forcing the atoms into a
smaller trapping volume - albeit at the cost of higher temperatures.  A density increase of one order of magnitude compared to a standard MOT has also been achieved in a `semi-dark' double MOT setup \cite{yang07}.  In dark spontaneous force optical trap (dark SPOT) configurations \cite{ketterle93, sinclair94,anderson94,kim01,townsend96}, cold atoms are allowed to
dissipate into an electronic state that does not couple to the trapping light, thereby shielding the atoms from
re-radiation and allowing atoms to accumulate at higher densities.

The ability to deliberately shape trapping or confining light allowed a variety of new geometries, including bottle beam traps \cite{li12}, holographic mesoscopic traps \cite{sebby-strabley05}  and blue detuned crossed dipole traps with a compressible dark core \cite{bienaime12}. While these traps achieve very high densities this is usually at the expense of high technical complexity, low atom numbers or heating due to compression.  In our experiment we combine holographic shaping of repump light with the mechanism of a dark SPOT to  produce a large trap at ultra-high density and without unwanted heating.

\begin{figure*}
\includegraphics[width=2\columnwidth]{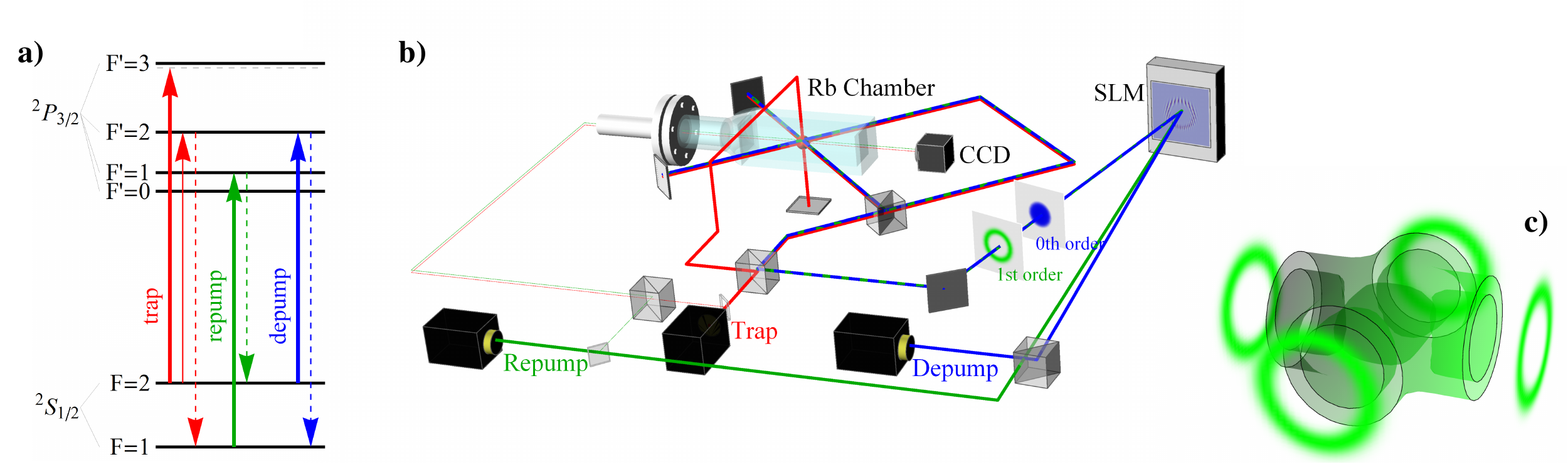}
\caption{\label{fig:setup}  a) Level scheme for $^{87}$Rb showing the frequencies of the laser beams (thick solid lines), off-resonant pumping (thin solid line) and relevant decay processes (dashed lines). b Simplified experimental setup; also showing the dark SPOT LG repump and depump profile. The beam expansion lenses have been omitted, all beams are collimated at the trap. c) Diagram of the repumping volume resulting from two crossed LG beams, created by drawing 2 isosurfaces.}
\end{figure*}

The original dark SPOT design recognised that capturing and cooling thermal atoms, and storing cold atoms, are best achieved in two separate regions, an outer shell and an inner core respectively. High intensity trapping light covers both regions.  In the outer shell, atoms that decay into untrapped states (in our case the $^{87}$Rb $F=1$ hyperfine state) are returned to the trapping cycle by a so-called `repump' laser, allowing efficent atom capture and cooling. In the central core, the repump laser is blocked so that cooled atoms can accumulate in the $F=1$ hyperfine state.  As this state does not interact with the intense trapping light, re-radiation is reduced and atoms can accumulate at far higher densities. In practice, simply blocking the repump light is not sufficient for rubidium and a further `depump' beam actively encourages the transfer to $F=1$ (see Fig.~\ref{fig:setup}a).    

Our dark SPOT system differs from all previously published implementations in two crucial ways: we shape the repump profile holographically rather than imaging an opaque circle into the trap center, and we switch dynamically between MOT and dark SPOT configurations. The experimental setup is shown schematically in Fig.~\ref{fig:setup}b and is in essence a standard MOT setup with an additional depump laser and a phase-only spatial light modulator (SLM, Hamamatsu LCOS). Any desired repump profile can be created by reflection off a suitably programmed SLM. The shaped repump light is crossed at the trap,  generating the shell-like repump volume illustrated in Fig.~\ref{fig:setup}c. At the same time we shape the depump laser complementary to the repump laser, directing depump intensity exactly to those areas where repump intensity is missing, namely the inner core of the trap, as indicated in the beam profiles incorporated in Fig.~\ref{fig:setup}b. 

We  initially capture atoms in a MOT with standard repump profile, in order to benefit from the large number of atoms in a (near) ideal MOT configuration.  By subsequently switching repump and depump lasers to the dark SPOT configuration, we transfer 75\% of these atoms into the dark SPOT, where they can accumulate at higher densities.

\begin{figure*}
\includegraphics[width=2\columnwidth]{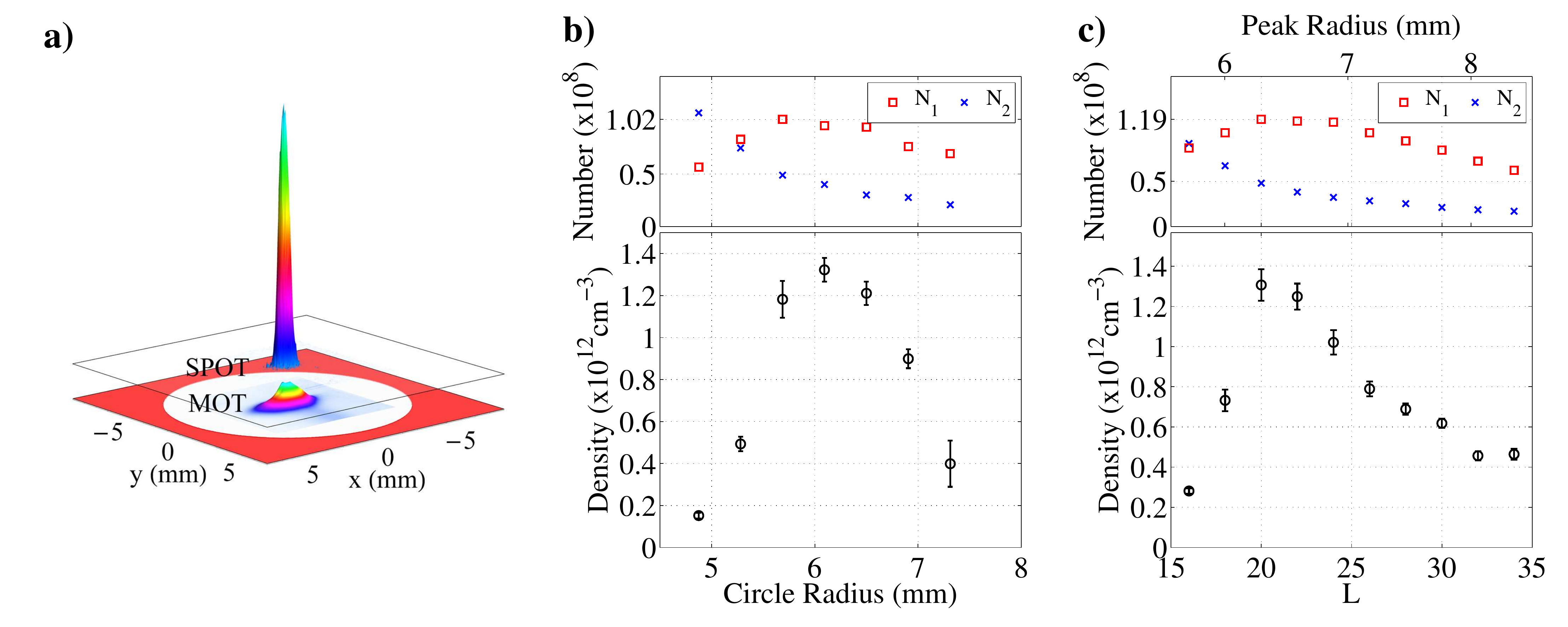}
\caption{\label{fig:MOTSPOT}  a) Characteristic two-dimensional atom densities measured for the MOT (below) and dark SPOT (above) configuration using a repump profile of a  6.5~mm radius dark disk beam, shown in red.  b) Dark SPOT densities and numbers as a function of the radius of the dark disk within the repump beam.  c) Dark SPOT densities and numbers for LG modes with equivalent Gaussian beam waist $w_0=2$~mm. In b) and c) the upper graph gives the number of atoms in the lower hyperfine ground state, $N_1$, and the upper hyperfine ground state $N_2$ while the lower graph shows the density of the atoms. The error bars in these plots are derived from the standard deviation taken from 10 measurements.}
\end{figure*}


\begin{figure}[!t]
\includegraphics[width=0.9\columnwidth]{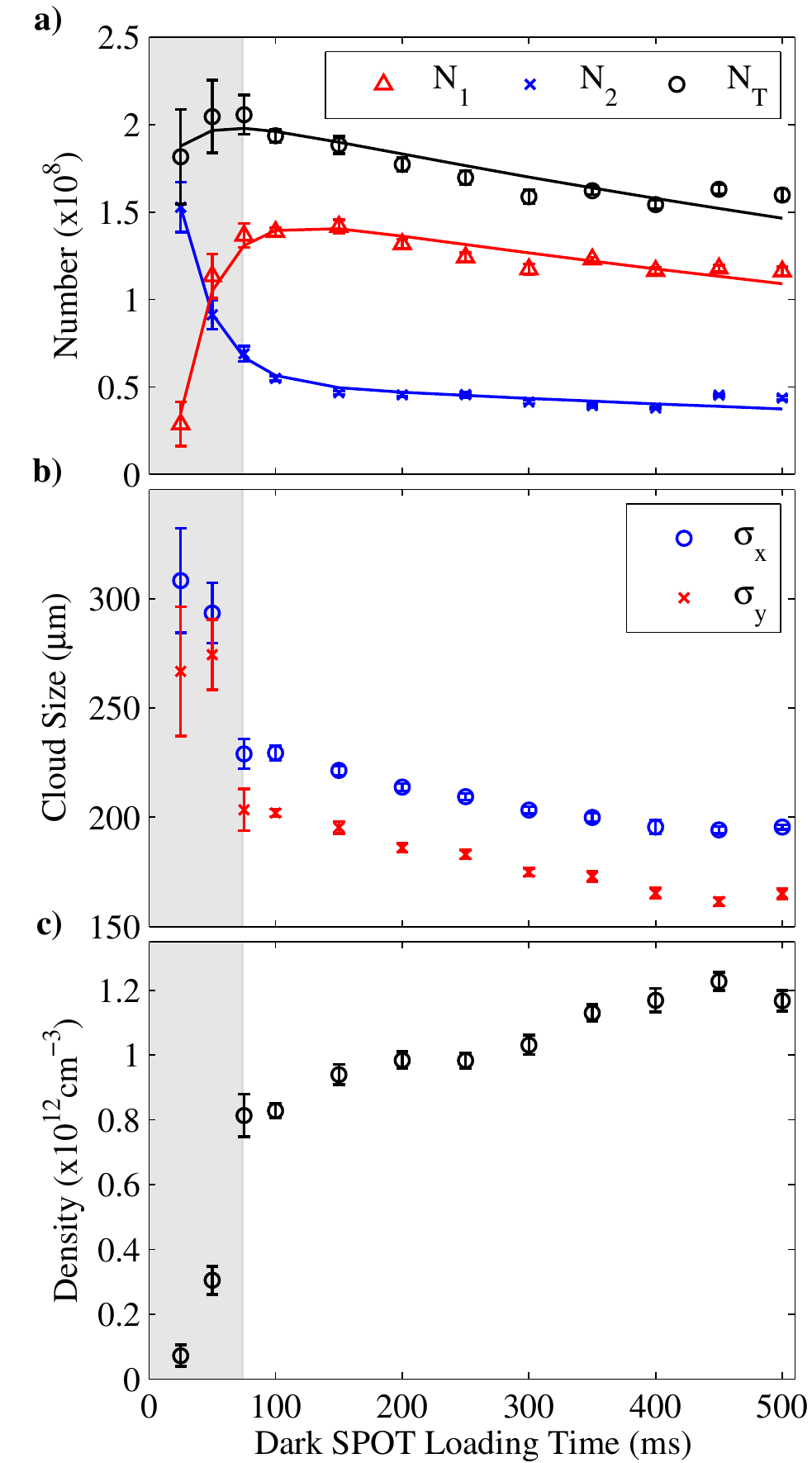}
\caption{\label{fig:DensityvsDepump} Number, size and density of the dark SPOT trap as a function of the time during which atoms are transferred from the MOT into the dark SPOT, for the example of a  6.5~mm radius dark disk repump. a) Number of atoms in the dark SPOT state $F=1$ ($N_1$), remaining atoms in the MOT state $F=2$ ($N_2$) and total atom number ($N_T$). The solid lines are fits to the data assuming a fast exponential decay of $N_2$  ($\tau=27$~ms) together with a slow exponential decay of $N_1$ ($\tau=1340$~ms). b) The width of the dark SPOT cloud in horizontal ($\sigma_x$) and vertical ($\sigma_y$) direction, where $\sigma$ is the standard deviation of the Gaussian distribution. c) Density of the dark SPOT. All error bars are derived from the standard deviation taken from 10 measurements.}
\end{figure}

\emph{Density variation with repump profile} - Here we discuss two different classes of repump profiles:  Laguerre-Gauss (LG) modes, that due to their phase singularity have a true intensity zero on axis \cite{allen03,andrews12};  and light modes with the intensity blocked from the center, `dark disk beams.' In both cases we have investigated the effect of changing the size of the dark core.  For the LG modes this is achieved by varying the winding number $l$ for a constant beam waist, generating a bright intensity ring at a radius proportional to $w_0 \sqrt{l/2}$.  

We start from a standard MOT with around $2 \times 10^{8}$ atoms at a peak density of $4 \times 10^{10}$~cm$^{-3}$, before switching to the dark SPOT repump configuration and simultaneously introducing a weak depump laser of typically 220~$\mu$W. The atoms evolve in this configuration for a depumping time of 250~ms before we determine the atom number and cloud size in the lower hyperfine level to obtain the dark SPOT density.

 In order to measure the density of atoms in a trap, one would typically perform a number measurement via absorption imaging and then either infer the size from the same measurement or via fluorescence imaging. However since absorption imaging relies on repeat absorption on a closed transition, this is not possible for atoms in the $F=1$ state, since no stretched state exists and on average each atom will only absorb 2 photons before being pumped into the  $F=2$ state. Therefore identifying the densities of atoms in the $F=1$ state requires measurements in three subsequent experimental realisations: (i) We determine the atom number, $N_2$, in $F=2$ by standard absorption imaging.  (ii) We repump atoms from $F=1$ into $F=2$ for 2~ms and measure the combined atom number, $N_{\rm T}$. This allows us to infer the number of atoms in the dark SPOT as $N_1=N_{\rm T}-N_2$. (iii) We measure the size of the $F=1$ cloud by weak absorption imaging on the $F=1$ to $F'=2$ transition, using the assumption that though the absorption is weak, the shape of the cloud is retained.  In combination with $N_1$  this yields the atom density in the lower hyperfine $F=1$ state.  

Independent of repump geometry we obtain the highest atom densities of $1.3 \times 10^{12}$~cm$^{-3}$ for a dark core radius of around 6~mm.  Surprisingly, the dark core region exceeds the size of the trapped atom cloud by far (6 times at e$^{-2}$ radius), and even more so compared to the dark SPOT cloud (20 times).    
A representative comparison of the different sizes is shown in  Fig.~\ref{fig:MOTSPOT}a for a typical dark disk beam, taken from the corresponding absorption and fluorescence measurements.  For the MOT and SPOT profiles the z axis is scaled to represent the correct ratio of 75\% of the MOT atoms  retained in the dark SPOT configuration.  The change of density and atom number as a function of the dark core radius can be seen in Fig.~\ref{fig:MOTSPOT}b and ~\ref{fig:MOTSPOT}c for dark disk and LG beams respectively.

Dark core regions of this size have never before been investigated. In previous dark SPOT experiments, atoms were trapped in situ, so that trap loading rates would be severely compromised for dark core regions exceeding or even close to the size of the MOT.  In our experiment we trap in a standard MOT before dynamically switching to the dark SPOT configuration, so that the dark SPOT geometry only needs to hold pre-cooled atoms. We reason that a large dark central core within the repump is essential in decreasing reradiation forces. The fraction of repump light that reaches the dark SPOT atoms from scattering in the repump region is proportional to $\sigma^2_{\rm dark SPOT}/ R^2$ where $\sigma_{\rm dark SPOT}$ is the cloud width and $R$ is the radius of the bright repump shell. Additionally, large dark beam regions reduce the amount of repump light that reaches the dark SPOT atoms directly via diffraction and scattering along the beam propagation. For dark core radii above 6~mm we observe a decrease in dark SPOT density. This may be explained by a reduction of the effective repump intensity which results from a decreasing overlap of the repump mode profiles with the repump beam incident on the SLM, in addition to unavoidable aperturing due to the beam hitting the edge of optical components along the repump path, resulting in beam size a limit of around 9~mm radius in our system.

\emph{Temporal Evolution} -   To investigate the loading and retention of atoms in the dark SPOT configuration we have monitored the number of atoms in both hyperfine ground states and the size of the dark SPOT cloud for a variety of depump times, see Fig.~\ref{fig:DensityvsDepump}. We identify two regimes:  a transient phase lasting for the first 75~ms, indicated by a grey shaded background, followed by a quasi steady-state phase. During the transient regime atoms are still being loaded from the trapped $F=2$ hyperfine state into $F=1$, identified by an increase of $N_1$ at the cost of $N_2$. The dark SPOT size decreases rapidly, resulting in an increase in density up to $8\times10^{11}$~cm$^{-3}$.  
After this initial phase the number of atoms in $F=2$ remains fairly constant, and we assume that atoms in $F=1$ slowly diffuse out of the dark SPOT region, re-enter the trapping cycle when reaching the repump shell, experience a force towards the trap center, and are pumped again into $F=1$ at the trap center. In this quasi steady-state regime, the dark SPOT density is still increasing, albeit at a slower timescale, suggesting that another mechanism has become dominant, one which develops over multiple cycles. The process seems limited only by the loss rate of atoms from the trap which results in a slow decrease of dark SPOT densities after the first 500~ms, until after 10~s there are essentially no atoms left (not shown).

\begin{figure}[t!]
\includegraphics[width=0.9\columnwidth]{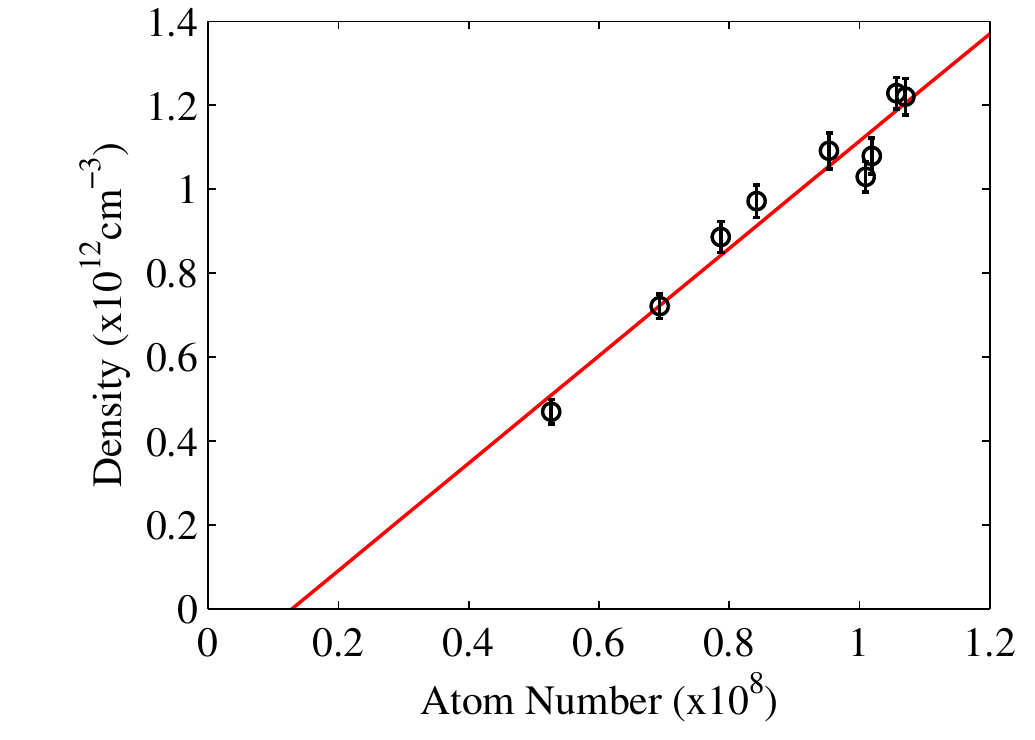}
\caption{\label{fig:MOTloading}  Scaling of dark SPOT density with initial MOT atom number. The solid line is a linear fit to the data. All data is taken after a dark SPOT loading time of 250~ms and for a repump with a dark disk radius of 6.5~mm. Error bars represent the standard deviation derived from 10 measurements.}
\end{figure}

\emph{Scalability with atom number} -  In order to test how our method might fare in systems with lower or higher initial atom numbers, we have artificially reduced the initial number of atoms to between  $5\times10^7$ and $1.1\times10^8$ atoms by limiting the MOT loading time before switching to the dark SPOT configuration. We found that the dark SPOT size remained stable within $10\%$ so that an increase in initial atom number translates directly into a linear increase in density, see Figure.~\ref{fig:MOTloading}. This indicates a robust mechanism that promises a successful extension to systems with both high and low atom numbers.

In contrast with compression type techniques for creating high density atomic clouds, our technique does not induce unwanted heating. We have verified this by determining the temperature via standard time of flight measurement performed by fluorescence imaging of the MOT and absorption imaging of the dark SPOT atoms respectively. For MOT and dark SPOT atoms we measured temperatures around $100~\mu K$, independent of depumping time and repump geometry.

\emph{Discussion} -  We have demonstrated extremely dense, long lived clouds of atoms by shaping the repump beam and introducing a depump beam. This is achieved in a technically simple system compared with systems involving molasses stages, compression, multiple MOTs or additional cooling techniques.  The atoms are shelved in a dark state which does not interact with any of the laser beams and accumulate in a narrow, dense cloud. We find that, independent of the specific repump geometry, a large repump core well beyond the atom trap dimension is required for the highest densities. We were able to  increase the density from $4\times10^{10}$ cm$^{-3}$ to $1.3\times10^{12}$ cm$^{-3}$, representing an improvement of almost two orders of magnitude in density at a stable temperature. As a consequence, the phase space density (PSD) increased also by almost two orders of magnitude up to $8.5\times10^{-6}$. We believe this to be a record PSD in a single cell trap of relatively low complexity, providing an ideal basis for experiments which require high PSD. The technique is stable, robust and potentially scalable to larger initial atom numbers, promising even higher atom densities.  It is expected that a similar mechanism could assist in the laser cooling of molecules \cite{demille10,ye13} and quantum degenerate gasses\cite{kuhr10,esslinger12}.

We acknowledge financial support by the European Commission via the FET Open grant agreement Phorbitech FP7-ICT-255914.

\end{document}